\documentclass[prd,tightenlines,12pt,showpacs]{revtex4}%
\usepackage{graphicx,color}
\usepackage{rotating}
\usepackage{amssymb}
\usepackage{color}
\usepackage{epsfig}
\usepackage{dcolumn}
\usepackage{bm}
\usepackage{multirow}
\usepackage{epstopdf}
\usepackage{amsmath}
\usepackage{amsfonts}
\usepackage{graphicx}%
\def\bes{\begin{subequations}}
\def\ees{\end{subequations}}

\def\be{\begin{equation}}
\def\ee{\end{equation}}
\def\ba{\begin{eqnarray}}
\def\ea{\end{eqnarray}}
\def\bear{\begin{array}}
\def\eear{\end{array}}

\begin{document}
\title{Heavy quark potential from QCD-related effective coupling }
\author{C\'esar Ayala, Pedro Gonz\'alez and Vicente Vento}
\affiliation{Department of Theoretical Physics and IFIC, University of Valencia and CSIC,
E-46100, Valencia, Spain}
\date{\today}

\begin{abstract}
We implement our past  investigations in the quark-antiquark interaction through a 
non-perturbative running coupling  defined  in terms of a gluon mass function, 
similar to that used in some Schwinger-Dyson approaches. This coupling leads 
to a quark-antiquark potential, which satisfies not only asymptotic freedom but 
also describes linear confinement correctly. From this potential, we calculate the 
bottomonium and charmonium spectra below the first open flavor meson-meson 
thresholds and show that for a small range of values of the free parameter 
determining the gluon mass function an excellent agreement with data is attained.
\end{abstract}

\maketitle

\section{Introduction}

The use of perturbation theory in quantum chromodynamics (QCD) \cite%
{Fritzsch:1973pi} provides a useful tool to study high energy processes as
confirmed by the accurate description of deep inelastic lepton-nucleon and
nucleon-nucleon collision processes \cite%
{Dokshitzer:1977sg,Gribov:1972ri,Altarelli:1977zs}.

Asymptotic freedom \cite{Gross:1973id,Politzer:1973fx} is the fundamental
property of QCD that justifies the use of perturbation methods to describe
the interaction among the constituents at high momentum transfers (short
distances).

In order to describe low energy (large distance) physics, for example the
hadron spectrum, perturbation theory cannot be used. The reason being that
the interaction among the constituents is dominated by another fundamental
property of QCD, the confinement of quarks and gluons \cite{Wilson:1974sk},
which is highly non-perturbative. Efforts have been made and techniques
developed to study this low momentum transfer regime, e.g. lattice QCD \cite{Bali:2000gf,Hagler:2011zz}, phenomenological quark potential models \cite{Eichten:1978tg,Quigg:1979vr,GI85}, studies on the nontrivial vacuum \cite{Kochelev:1985de}, effective theories \cite{Brambilla:1999ja}, dispersive
extensions of the beta function \cite{Nesterenko:2001st,Nesterenko:2003xb}, and the 
solution of the Schwinger-Dyson equations \cite{Cornwall:1982zr,Aguilar:2006gr}. {Another development which has led to very succesful results in the study of the spectrum and the behavior of the strong coupling in the infrared (IR) is AdS/CFT \cite{Erlich:2005qh,Brodsky:2014yha}.}

The resolution of Schwinger-Dyson equations (SDE)  leads to the freezing
of the QCD running coupling (effective charge) in the infrared which can be conveniently
understood as a dynamical generation of a gluon mass function, giving rise to a momentum dependence which 
is free from IR divergences \cite{Cornwall:1982zr,Aguilar:2006gr, Aguilar:2009nf, Ayala:2012pb, Aguilar:2013hoa} . It was shown that the 
interquark static potential for heavy mesons described by a massive one gluon exchange interaction 
obtained from the propagator of the truncated Schwinger-Dyson equations does not reproduce the 
expected Cornell potential \cite{Gonzalez:2011zc}. This was attributed to the lack of a mechanisms 
for confinement \cite{Gonzalez:2011zc} by explicit  comparison with lattice QCD calculations 
\cite{Greensite:2003xf}.

{To solve this problem an {\it ad hoc} effective mechanism was proposed based on a singular nonperturbative coupling 
 \cite{Vento:2012wp}, which generates a Gribov singularity in the potential while keeping the propagator 
infrared finite. We  show here that the proposal leads to a potential  with linear  confinement and the correct 
asymptotic behavior of QCD.} The resulting potential is Cornell like for intermediate distances ($0.1-4$ fm) and 
for a small range of values of the free parameter determining the gluon mass function it provides a good spectral 
description of heavy quarkonia.

{Other authors have also used non-perturbative effective couplings which reproduce asymptotic freedom 
in the UV and have $1/q^2$ behavior in the IR to study the spectrum and  other physical observables 
\cite{Richardson:1978bt, Buchmuller:1980su,Nesterenko:1999np,Epele:2001ic,Luna:2010tp,Gomez:2015tqa}.
These couplings have been justified from QCD under certain requirements 
\cite{Gribov:1999ui,Simonov:1999qj,Yndurain:2000yq}.}
{However, it must be noted that many succesful phenomenological studies of the IR coupling 
point to a finite value \cite{Deur:2016tte}.}

The contents of this paper are organized as follows.
In Section \ref{SII} we establish the prescription to get the static
quark-antiquark potential from the running coupling.{ A possibility  to generate a
linear confining potential is through a strong Gribov singularity for small momenta.} We propose a 
functional form resulting from a modification of 
the Schwinger-Dyson solution for the coupling, which has the {desired} 
singularity structure at small momenta and satisfies asymptotic freedom 
at large momenta. The presence of the singularity requires, in order to get 
the potential, a regularization scheme which is described in detail in Section \ref{SIII}.
Then, in Section \ref{SIV}, the results obtained are presented. The
potential is compared to the Cornell one. The calculated heavy quarkonia
masses are compared with data and with the ones resulting from
\textquotedblleft equivalent\textquotedblright\ Cornell potentials. Finally,
in Section \ref{SV}, the main conclusions of this work are summarized.

\section{Quark-Antiquark Potential from the non-perturbative running coupling
\label{SII}}

The history of the theoretical description of the quark-antiquark potential
and its relation to QCD has been motivated by the tremendous
phenomenological success of the Cornell potential \cite%
{Eichten:1978tg,Quigg:1979vr} 
\begin{equation}
V^{C}(r)=-\frac{\chi }{r}+\sigma r\ ,  \label{Cornellpot}
\end{equation}%
where $\chi $, the Coulomb strength, and $\sigma $, the string tension, are
constants to be fitted from data. It is amazing that this simple potential
reproduces quite accurately the experimental heavy quark meson spectra below
the open flavor meson-meson threshold energies (see Section \ref{SIV}).

Two limits characterize this potential. The short distance limit which
satisfies asymptotic freedom, and the long distance limit which describes the
IR behavior, i.e. confinement of heavy static sources.

In general, a quark-antiquark potential in configuration space, $V(r),$
reads (note that the angular variables have been integrated)

\begin{equation}
V(r)=-\frac{2}{\pi }\int_{0}^{\infty }V(q^{2})q^{2}\frac{\sin (qr)}{qr}dq%
\text{ ,}
\label{fourier}
\end{equation}%
where $V(q^{2})$ is the potential in momentum space. 

By assuming one gluon exchange dominance, $V(q^{2})$ can be derived from QCD
as

\begin{equation}
V(q^{2})=\frac{4}{3}\alpha _{s}(q^{2})\Delta (q^{2})  \label{potential}
\end{equation}%
with $\alpha _{s}(q^{2})$ and $\Delta (q^{2})$ standing for the running
coupling and the gluon propagator respectively.

It is now well established from lattice QCD, and confirmed by
Schwinger-Dyson calculations, that the finite character of the propagator (at $%
q^{2}\rightarrow 0$) reads \cite{Cornwall:1982zr,Aguilar:2006gr}%
\begin{equation}
\Delta (q^{2})=\frac{1}{q^{2}+m_{g}^{2}(q^{2})}  \label{propq}
\end{equation}%
where $m_{g}(q^{2})$ is an effective gluon mass (see below). {Other points of view to parametrize
the gluon propagator have been developed, like flux tubes \cite{GI85} and effective potentials \cite{Brodsky:2014yha}, 
in which the connection to the concept of massive gluon exchange is not currently known.}

Regarding the coupling, a truncated solution of a gauge invariant subset of
the Schwinger-Dyson equations for QCD gives rise to a functional form
showing a freezing in the IR  {which can be cast as \cite{Aguilar:2006gr} : }

\begin{equation}
\alpha _{s}^{\mathrm{(SD)}}(q^{2})=\frac{4\pi }{\beta _{0}\mathrm{ln}\left( 
\frac{q^{2}+m_{g}^{2}(q^{2})}{\Lambda ^{2}}\right) }  \label{DScoupling}
\end{equation}%
where $\beta _{0}=11-2/3\ n_{f}$ is the first $\beta $-function coefficient
for QCD, $n_{f}$ being the number of active quarks, $\Lambda $ is the scale
parameter in QCD and $m_{g}^{2}(q^{2})$ is a gluon mass given by \cite{Aguilar:2014tka} 

\begin{equation}
m_{g}^{2}(q^{2})=\frac{m_{0}^{2}}{1+(q^{2}/\mathcal{M}^{2})^{1+p}}\ ,
\label{gmass}
\end{equation}%
with $m_{0}=m_{g}(q^{2}=0)$, $\mathcal{M}$ and $p>0$ constants. 
{Other parametrizations for both coupling and gluon mass can be found in ref. \cite{Deur:2016tte}.}
The coupling (\ref{DScoupling}) is to first order the one proposed in  \cite{Aguilar:2009nf} 
obtained from the gluon propagator using the pinch technique.
{Note that the dynamical gluon mass varies with the number of flavors ($n_f$) \cite{Ayala:2012pb,Aguilar:2013hoa}
and therefore is determined by the initial $m_0$ and the QCD scale $\Lambda$.}

The combination (\ref{potential}), (\ref{propq}), (\ref{DScoupling}) and 
(\ref{gmass})  is renormalization group invariant since the ghost contribution 
has been incorporated into the definition of the coupling 
 \cite{Gonzalez:2011zc}. Therefore one can construct
a potential. The result is a non linearly rising potential \cite%
{Gonzalez:2011zc} contrary to the expectation from the quenched
approximation followed to derive the propagator and the coupling. Actually a
linearly rising Cornell-like behavior has been obtained in quenched lattice
QCD \cite{Bali:2000gf,Greensite:2003xf}. One possible reason for this
anomaly might be that confinement is more than a one gluon exchange effect.
An alternative possibility, suggested in reference \cite{Vento:2012wp}, is
that the Schwinger-Dyson coupling contains the physics at intermediate and
large $q^{2}$ but it lacks some vertex corrections at low $q^{2}$. We shall examine 
this alternative next.

For this purpose we realize
that in order to lead to linear confinement the mass singularity of the
propagator must be eliminated and instead a Gribov singularity should appear. An
easy way to do this from (\ref{DScoupling}) is through the modified coupling%

\begin{equation}
\alpha _{s}(q^{2})\equiv \left( \alpha _{s}^{\mathrm{(SD)}}(q^{2})\right)
_{m_{0}=\Lambda }\frac{\left( q^{2}+m_{g}^{2}(q^{2})\right) }{q^{2}}
\label{modDS}
\end{equation}%
so that the factor $\left( q^{2}+m_{g}^{2}(q^{2})\right) $ cancels the
singularity of the propagator whereas the factor $\frac{1}{q^{2}}$ gives
rise altogether with $\left( \alpha _{s}^{\mathrm{(SD)}}(q^{2})\right)
_{m_{0}=\Lambda }$ to a $\frac{1}{q^{4}}$ Gribov singularity. In this way
the low $q^{2}$ behavior of $\alpha _{s}^{\mathrm{(SD)}}(q^{2})$ has been
corrected as required while the intermediate and long distance $q^{2}$
dependencies are preserved since $m_{g}^{2}$ is a quickly decreasing function
with $q^{2}.$ {When $m_0 = \Lambda$ our potential
becomes in the IR limit, $q^{2}<<\Lambda^{2}$, that of Richardson \cite{Richardson:1978bt}, 
signaling that it is defined in the $\overline{MS}$ scheme.}
Moreover, $\alpha _{s}(q^{2})$ reproduces in the
asymptotic limit $q^{2}>>\Lambda ^{2}$ the well known one-loop perturbative
QCD result 

\begin{equation*}
\alpha _{s}^{pert}(q^{2})=4\pi /(\beta _{0}\mathrm{ln}(q^{2}/\Lambda ^{2}))%
\text{ }\ ,
\end{equation*}%
Additional terms of the order $\sim\mathcal{O}(1/(\mathrm{ln}(q^{2}/\Lambda ^{2}) q^N))$, where $N=4+2p>4$, 
have to be added to this result but hey are negiglible and arise from the IR behavior \cite{ITEP}.

In order to check whether this modified coupling proposal is physically
meaningful or not we shall study, from the potential deriving from it, heavy
quarkonia, using as a criterion of meaningfulness the accurate description
of the spectra. More precisely, we shall choose for the evolution mass
parameters $\mathcal{M}$ and $p$ their estimated values in the Schwinger 
Dyson context \cite{Aguilar:2014tka} 
\begin{equation*}
\mathcal{M}=436\text{ MeV}\mbox{ and }p=0.15,
\end{equation*}%
while leaving $\Lambda$ ($=m_0$) as the $only$ free parameter of the potential.
Then, an accurate spectral description for a reasonable value of $\Lambda$ 
might be {an indication that the proposed coupling parametrizes efficiently 
and conveniently the phenomenology.}

It must be stressed at this point that we have incorporated the Gribov singularity 
guided by lattice QCD and simplicity and in so doing we have departed from the 
SDE approach. Nevertheless the effective way of introducing the coupling mantains 
the behavior of the SDE result for large and intermediate energies and therefore 
a good fit of the spectrum is strong support of the SDE calculation.

\section{Regularization procedure\label{SIII}}

From  (\ref{potential}), (\ref{propq}) and  (\ref{modDS}) the potential reads%

\begin{equation}
V(r)=-\frac{32}{3\beta_{0}}\int_{0}^{\infty}\frac{1}{\mathrm{ln}\left(
\frac{q^{2}+\frac{\Lambda^{2}}{1+(q^{2}/\mathcal{M}^{2})^{1+p}}}{\Lambda^{2}%
}\right)  }\frac{\sin(qr)}{qr}dq\text{ .} \label{potvv}%
\end{equation}

The integral in (\ref{potvv}) is divergent since the integrand behaves as
$\frac{\Lambda^{2}}{q^{2}}$ at the origin $q\mapsto0$. To extract its physical
content we have to regularize it. To understand how the regularization
procedure works let us consider the simpler case of the potential (with the
same singular behavior at the origin)%
\[
\widetilde{V}(q^{2})=\frac{4\nu}{3}\frac{\Lambda^{2}}{q^{4}}%
\]
where $\nu$ is an dimensionless constant. Its Fourier transform, from
(\ref{fourier}), is given by%
\[
V^{R}(r)=-\frac{8\nu\Lambda^{2}}{3\pi}\int_{0}^{\infty}\frac{1}{q^{2}}%
\frac{\sin(qr)}{qr}dq
\]
By introducing a regulator $\lambda$ we can rewrite it as%
\begin{align*}
V^{R}(r)  &  =-\frac{8\nu\Lambda^{2}}{3\pi}\frac{1}{r}\lim_{\lambda
\rightarrow0}\int_{0}^{\infty}\frac{q}{\left(  q^{2}+\lambda^{2}\right)  ^{2}%
}\sin(qr)dq\\
&  =-\frac{8\nu\Lambda^{2}}{3\pi}\frac{1}{r}\lim_{\lambda\rightarrow0}\left(
\operatorname{Im}\frac{1}{2}\int_{-\infty}^{\infty}\frac{q}{\left(
q^{2}+\lambda^{2}\right)  ^{2}}e^{iqr}dq\right)
\end{align*}
so that the integral can be solved analytically in the complex plane giving
\[
V^{R}(r)=\frac{2\nu\Lambda^{2}}{3}\lim_{\lambda\rightarrow0}\left(  -\frac
{1}{\lambda}\right)  +\frac{2\nu\Lambda^{2}}{3}r
\]

The divergent term $\frac{2\nu\Lambda^{2}}{3}\lim_{\lambda\rightarrow0}\left(
-\frac{1}{\lambda}\right)  $ does not depend on $r$. As the potential is
defined up to an arbitrary constant we may simply remove such spurious
divergence from the potential. To do this in a more systematic way, let us
realize that the divergent term comes out from the behavior of the integrand
at $q^{2}\mapsto0$. Then, by making an expansion of the integrand around
$q=0$,
\[
\frac{1}{q^{2}}\frac{\sin(qr)}{qr}=\frac{1}{q^{2}}-\frac{(qr)^{3}}{6q^{2}%
}+...\text{ ,}%
\]
we can easily identify the first term $\frac{1}{q^{2}}$ as causing the
divergence of the integral at $q^{2}\mapsto0$. Indeed, if we integrate this
term with the same regularization procedure we get%
\[
-\frac{8\nu\Lambda^{2}}{3\pi}\int_{0}^{\infty}\frac{q^{2}}{\left(
q^{2}+\lambda^{2}\right)  ^{2}}dq=-\frac{2\nu\Lambda^{2}}{3\lambda}%
\]

Therefore we can redefine the potential by subtracting this non physical
divergence as%
\[
V(r)\equiv-\frac{8\nu\Lambda^{2}}{3\pi}\lim_{\lambda\rightarrow0}\left(
\int_{0}^{\infty}\frac{q^{2}}{\left(  q^{2}+\lambda^{2}\right)  ^{2}}%
\frac{\sin(qr)}{qr}dq-\int_{0}^{\infty}\frac{q^{2}}{\left(  q^{2}+\lambda
^{2}\right)  ^{2}}dq\right)
\]

Back to (\ref{potvv}) it is more convenient to regularize it through a cutoff
$\gamma$ in the form%
\[
V(r)=-\frac{32}{3\beta_{0}}\lim_{\gamma\rightarrow0}\int_{\gamma}^{\infty
}\frac{1}{\mathrm{ln}\left(  \frac{q^{2}+\frac{\Lambda^{2}}{1+(q^{2}%
/\mathcal{M}^{2})^{1+p}}}{\Lambda^{2}}\right)  }\frac{\sin(qr)}{qr}dq
\]
To subtract the spurious divergence we expand the integrand around $q=0:$
\begin{align*}
&  \frac{1}{\mathrm{ln}\left(  \frac{q^{2}+\frac{\Lambda^{2}}{1+(q^{2}%
/\mathcal{M}^{2})^{1+p}}}{\Lambda^{2}}\right)  }\frac{\sin(qr)}{qr}\\
&  =\frac{\Lambda^{2}}{q^{2}}\left(  1+\frac{\Lambda^{2}}{\mathcal{M}^{2+2p}%
}q^{2p}+\frac{\Lambda^{4}}{\mathcal{M}^{4+4p}}q^{4p}+\frac{\Lambda^{6}%
}{\mathcal{M}^{6+6p}}q^{6p}+...\right)
\end{align*}
and keep only the terms giving rise after integration to a singular behavior
at $q^{2}\mapsto0$ (they correspond, for $p=0.15$, to the explicitly written
ones). By integrating these terms with the chosen cutoff regularization we get%
\begin{align*}
\frac{I_{s}\left(  \gamma\right)  }{\Lambda^{2}}  &  \equiv\int_{\gamma
}^{\infty}\frac{1}{q^{2}}\left(  1+\frac{\Lambda^{2}}{\mathcal{M}^{2+2p}%
}q^{2p}+\frac{\Lambda^{4}}{\mathcal{M}^{4+4p}}q^{4p}+\frac{\Lambda^{6}%
}{\mathcal{M}^{6+6p}}q^{6p}\right)  dq\\
&  =\frac{1}{\gamma}+\frac{\Lambda^{2}}{\mathcal{M}^{2+2p}}\frac{1}{\left(
1-2p\right)  }\frac{1}{\gamma^{1-2p}}+\frac{\Lambda^{4}}{\mathcal{M}^{4+4p}%
}\frac{1}{\left(  1-4p\right)  }\frac{1}{\gamma^{1-4p}}\\
&  +\frac{\Lambda^{6}}{\mathcal{M}^{6+6p}}\frac{1}{\left(  1-6p\right)  }%
\frac{1}{\gamma^{1-6p}}%
\end{align*}
so that the physical potential reads%
\begin{equation}
V(r)=-\frac{32}{3\beta_{0}}\lim_{\gamma\rightarrow0}\left(  \int_{\gamma
}^{\infty}\frac{1}{\mathrm{ln}\left(  \frac{q^{2}+\frac{\Lambda^{2}}%
{1+(q^{2}/\mathcal{M}^{2})^{1+p}}}{\Lambda^{2}}\right)  }\frac{\sin(qr)}%
{qr}dq-I_{s}\left(  \gamma\right)  \right)  \label{finalpotvv}%
\end{equation}
which is evaluated numerically.

\section{Results\label{SIV}}

\begin{figure}[ptb]
\begin{center}
\includegraphics[
height=2.2739in,
width=3.6207in]%
{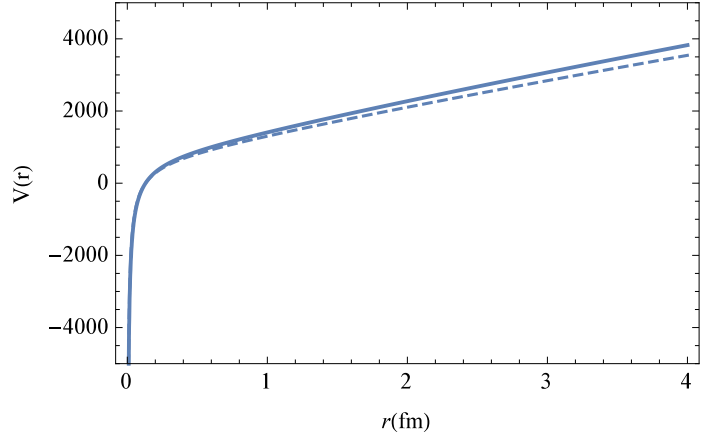}%
\caption{Bottomonium (solid) and charmonium (dashed) potentials for
$\Lambda=320$ MeV.}%
\label{Figpotbc320}%
\end{center}
\end{figure}

In order to fix $\Lambda$, the only free parameter of the potential, we
require (\ref{finalpotvv}) to provide a reasonable description of the heavy
quarkonia (bottomonium and charmonium) spectra. As will be justified later on, 
for bottomonium, one has to use
$n_{f}=4$ and for charmonium $n_{f}=3$. To
calculate the spectrum we solve the Schr\"{o}dinger equation. For any value of
$\Lambda$, we choose the quark masses, $m_{b}$ and $m_{c}$, to get the best
spectral fit. In this regard, as (\ref{finalpotvv}) represents a quenched
potential we restrict the comparison with data to energies below the
corresponding open flavor meson-meson thresholds. 

It turns out that only for a quite restricted range of values of $\Lambda$
 ($\Lambda_c \sim \Lambda_b \sim 320$ MeV) a good spectral description for bottomonium and
charmonium is obtained. The corresponding potentials for $\Lambda=320$ MeV are
shown in Fig. \ref{Figpotbc320}.

As can be seen, the potential (\ref{finalpotvv}) shows a soft flavor dependence
in the slope for intermediate and large distances. However, if we use $n_f=3$ also for bottomonium
the calculated spectral masses will only change slightly.

In Tables~\ref{tablespectrumb} and \ref{tablespectrumc} we list the calculated
masses for bottomonium and charmonium for $\Lambda=320$ MeV as compared to
data. To denote the states we use the spectroscopic notation $nl,$ in terms of
the radial, $n,$ and orbital angular momentum, $l,$ quantum numbers of the
quark-antiquark system. As we are dealing with a spin-independent potential we
compare as usual the calculated $s-$wave state masses with spin-triplet data,
the $p-$wave state masses with the centroids obtained from data and the $d-$wave 
states with the few existing experimental candidates.


\begin{table}[ptb]%
\begin{tabular}
[c]{ccccc}%
$J^{PC}$ & $nl$ & $%
\begin{array}
[c]{c}%
M_{V(r)_{\Lambda=320\text{ MeV}}}\\
\text{MeV}%
\end{array}
$ & $%
\begin{array}
[c]{c}%
M_{PDG}\\
\text{MeV}%
\end{array}
$ & $%
\begin{array}
[c]{c}%
M_{V^{C}(r)_{b\overline{b}}}\\
\text{MeV}%
\end{array}
$\\\hline
&  &  &  & \\
$1^{--}$ & $1s$ & $9489$ & $9460.30\pm0.26$ & $9479$\\
& $2s$ & $10023$ & $10023.26\pm0.31$ & $10013$\\
& $1d$ & $10147$ & $10163.7\pm1.4$ & $10155$\\
& $3s$ & $10354$ & $10355.2\pm0.5$ & $10339$\\
& $2d$ & $10435$ &  & $10427$\\
& $4s$ & $10621$ & $10579.4\pm1.2$ & $10596$\\
& $3d$ & $10681$ &  & $10666$\\
& $5s$ & $10854$ & $10876\pm11$ & $10825$\\
&  &  &  & \\
& $4d$ & $10903$ &  & $10883$\\
&  &  &  & \\
&  &  &  & \\
&  &  &  & \\
$\left(  0,1,2\right)  ^{++}$ & $1p$ & $9903$ & $9899.87\pm0.28\pm0.31$ &
$9920$\\
&  &  &  & \\
$\left(  0,1,2\right)  ^{++}$ & $2p$ & $10254$ & $10260.24\pm0.24\pm0.50$ &
$10252$\\
&  &  &  & \\
$\left(  0,1,2\right)  ^{++}$ & $3p$ & $10531$ & $10534\pm9$ & $10519$\\
&  &  &  &
\end{tabular}
\caption{Calculated $J^{PC}$ bottomonium masses from $V(r)_{\Lambda=320\text{
MeV}}$ and $m_{b}=4450$ MeV$.$ Masses for experimental resonances, $M_{PDG},$
have been taken from \cite{PDG14}. For $1p$ and $2p$ states the experimental
centroids are quoted. For $3p$ states the only known experimental mass is
listed. Masses from $V^{C}(r)_{b\overline{b}}$ , the \textquotedblleft
equivalent\textquotedblright\ Cornell potential (\ref{eqcornellb}) and the
same quark mass are also shown for comparison.}%
\label{tablespectrumb}%
\end{table}


\begin{table}[ptb]%
\begin{tabular}
[c]{ccccc}%
$J^{PC}$ & $nl$ & $%
\begin{array}
[c]{c}%
M_{V(r)_{\Lambda=320\text{ MeV}}}\\
\text{MeV}%
\end{array}
$ & $%
\begin{array}
[c]{c}%
M_{PDG}\\
\text{MeV}%
\end{array}
$ & $%
\begin{array}
[c]{c}%
M_{V^{C}(r)_{c\overline{c}}}\\
\text{MeV}%
\end{array}
$\\\hline
&  &  &  & \\
$1^{--}$ & $1s$ & $3090$ & $3096.916\pm0.011$ & $3105$\\
& $2s$ & $3671$ & $3686.108_{-0.014}^{+0.011}$ & $3671$\\
& $1d$ & $3772$ & $3778.1\pm1.2$ & $3772$\\
& $3s$ & $4102$ & $4039\pm1$ & $4096$\\
& $2d$ & $4170$ & $4191\pm5$ & $4166$\\
&  &  &  & \\
&  &  &  & \\
$\left(  0,1,2\right)  ^{++}$ & $1p$ & $3484$ & $3525.30\pm0.11$ & $3493$\\
&  &  &  & \\
$2^{++}$ & $2p$ & $3940$ & $3927.2\pm2.6$ & $3940$\\
&  &  &  &
\end{tabular}
\caption{Calculated $J^{PC}$ charmonium masses from $V(r)_{\Lambda=320\text{
MeV}}$ and $m_{c}=1030$ MeV$.$ Masses for experimental resonances, $M_{PDG},$
have been taken from \cite{PDG14}. For $1p$ states the experimental centroid
is quoted. For $2p$ states we quote the $2^{++}$ state which lies below the
$2^{++}$ threshold. Masses from $V^{C}(r)_{c\overline{c}}$ , the
\textquotedblleft equivalent\textquotedblright\ Cornell potential
(\ref{eqcornellc}), and the same quark mass are also shown for comparison.}%
\label{tablespectrumc}%
\end{table}
A very good spectral description is attained. Note that the only
significant difference ($>60$ MeV) between the calculated masses and data is
for the $3s$ charmonium state and it may be explained through configuration
mixing with the $2d$ one.

For further comparison the spectra from \textquotedblleft
equivalent\textquotedblright\ Cornell potentials have also been quoted. This
equivalence is based on the observation that for intermediate distances
($0.1-4$ fm) the potentials (\ref{finalpotvv}) can be very well approximated
by the Cornell types
\begin{equation}
\left(  V^{C}(r)\right)  _{b\overline{b}}=\sigma_{b\overline{b}}r-\frac
{\chi_{b\overline{b}}}{r}+a_{b\overline{b}}\label{eqcornellb}%
\end{equation}
with%
\begin{align}
\sigma_{b\overline{b}} &  =800\text{ MeV.fm}^{-1},\label{parcornellb}\\
\chi_{b\overline{b}} &  =100\text{ MeV.fm},\nonumber\\
a_{b\overline{b}} &  =693\text{ MeV},\nonumber
\end{align}
and%
\begin{equation}
\left(  V^{C}(r)\right)  _{c\overline{c}}=\sigma_{c\overline{c}}r-\frac
{\chi_{c\overline{c}}}{r}+a_{c\overline{c}}\label{eqcornellc}%
\end{equation}
with%
\begin{align}
\sigma_{c\overline{c}} &  =735\text{ MeV.fm}^{-1},\label{parcornellc}\\
\chi_{c\overline{c}} &  =100\text{ MeV.fm},\nonumber\\
a_{c\overline{c}} &  =658\text{ MeV},\nonumber
\end{align}
where the fitted Coulomb strengths are in agreement with the values derived
from QCD from the hyperfine splitting of $1p$ states in bottomonium
\cite{Ynd95} and from the fine structure splitting of $1p$ states in
charmonium \cite{Bad99}.

\begin{figure}
[ptb]
\begin{center}
\includegraphics[
height=2.3163in,
width=3.6207in
]%
{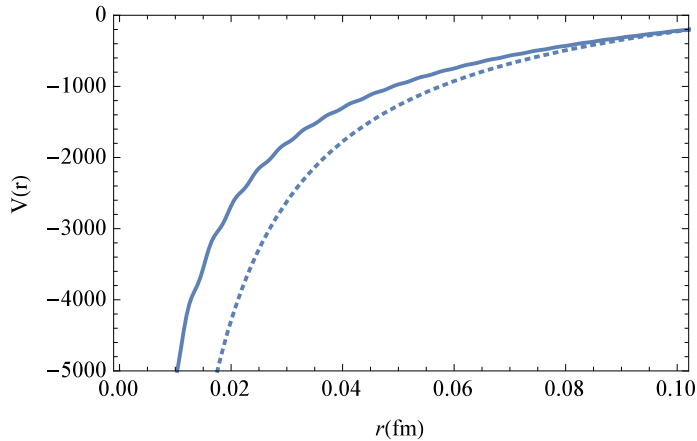}%
\caption{Bottomonium potential for $\Lambda=320$ MeV (solid line) versus its
\textquotedblleft equivalent\textquotedblright\ Cornell potential (dotted
line) for short distances.}%
\label{Figshort320}%
\end{center}
\end{figure}

\begin{figure}
[ptbptb]
\begin{center}
\includegraphics[
height=2.344in,
width=3.6207in
]%
{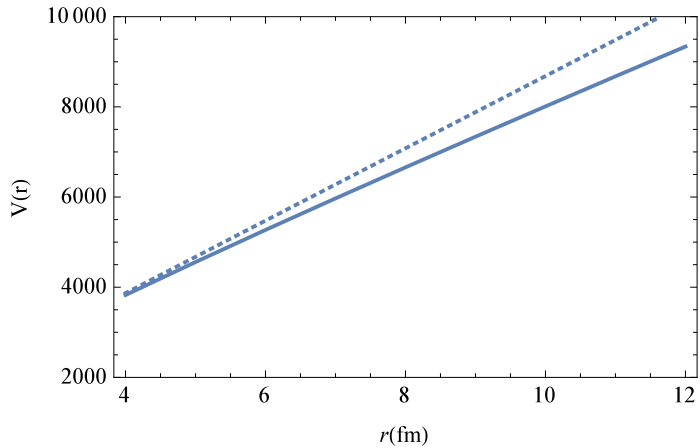}%
\caption{Bottomonium potential for $\Lambda=320$ MeV (solid line) versus its
\textquotedblleft equivalent\textquotedblright\ Cornell potential (dotted
line) for long distances.}%
\label{Figlong320}%
\end{center}
\end{figure}

As a matter of fact, in the mentioned region, they can not be distinguished from those in Fig.
\ref{Figpotbc320}. However, below and above this region they
become  different as shown in Figs. \ref{Figshort320} and
\ref{Figlong320} for bottomonium.

These differences have not much effect on the calculated masses, as can be
seen in Tables~\ref{tablespectrumb} and \ref{tablespectrumc}. However, in
order to get a more accurate fit to the spectra from a Cornell potential the
string tensions have to be slightly increased with respect to those of
(\ref{eqcornellb}) and (\ref{eqcornellc}).

It is interesting to check {\it a posteriori} our initial assumption about the
number of active quarks $n_{f}$ ($3$ for charmonium and $4$ for bottomonium).
The momentum determines the active number of flavors in the coupling, thus if
\[m_{f}^{2}c^{2}<q^{2}=\left(  \overrightarrow{p}_{q}-\overrightarrow
{p}_{\overline{q}}\right)  ^{2}=4\left\vert \overrightarrow{p}_{q}\right\vert
^{2}%
\]
where $\overrightarrow{p}_{q}$ is the three-momentum of the quark (charm or
bottom) in the center-of-mass system, then  $n_f$  is to be used in the coupling. 
This can be conveniently rewritten as
\[
m_{f}^{2}c^2<m_{q}^{2}v^{2}%
\]
where $v$ is the relative quark-antiquark velocity. By using the values
calculated from the $1s$ wave functions, $\left(  v_{1s}^{2}\right)
_{c\overline{c}}=0.3c^{2}$ and  $\left(  v_{1s}^{2}\right)  _{b\overline{b}%
}=0.09c^{2},$ we can see that in charmonium $q > m_s c$, $s$ for strangeness, and therefore $n_{f}=3$, while
in bottomonium, $q > m_c c$ and therefore $n_{f}=4$.

It is  illustrative to compare the potential (\ref{finalpotvv}) fitting
the charmonium spectrum (with $\Lambda=320$ MeV, see Table~\ref{tablespectrumc})
with the potential of Richardson \cite{Richardson:1978bt} obtained from (\ref{potential}) 
with $\Delta(q^{2})\equiv\frac{1}{q^{2}}$ and 

\begin{equation}
\alpha_{s}^{\mathrm{(R)}}(q^{2})=\frac{4\pi}{\beta_{0}\mathrm{ln}\left(
1+q^{2}/\Lambda^{2}\right)  }\ . \label{alphaRichardson}%
\end{equation}

This comparison is shown in Fig. \ref{FigcompRichardson}. Although generated from different 
approaches, both potentials give a similar quality fits to the spectrum. 
{ {It must be recalled that both potentials are defined in the $\overline{MS}$ scheme. 
The value of $\Lambda$ required by equation (\ref{finalpotvv}) ($\Lambda= 320$ MeV)
is in better agreement  with QCD than Richardson's ($\Lambda = 398$ MeV)  \cite{Richardson:1978bt}, 
since in the  $\overline{MS}$ scheme, $\Lambda$ (4 Loops, $n_f =  3$) $= 358 \pm 22$ MeV and $\Lambda$ 
(4 Loops, $n_f =  4$) $= 303 \pm 21$ MeV 
\cite{Yndurain:1999ui,Kataev:2001kk,Kataev:2015yha}.}

\begin{figure}
[ptb]
\begin{center}
\includegraphics[
height=2.2739in,
width=3.6207in
]%
{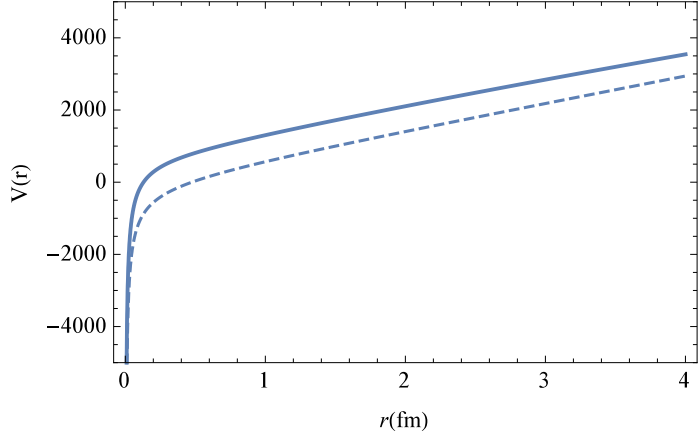}%
\caption{Charmonium potential (solid line) for $\Lambda=320$ MeV ($n_{f}=3$) as compared to
Richardson's potential (dashed line) for $\Lambda=398$ MeV ($n_{f}=3$).}%
\label{FigcompRichardson}%
\end{center}
\end{figure}

\section{Conclusions\label{SV}}

This work has been motivated by previous studies aiming at describing the phenomenological successful potentials from QCD \cite{Gonzalez:2011zc,Vento:2012wp}. Our aim here  has been to describe the heavy quark spectra from  a potential derivable from non-perturbative QCD studies. Our research has led to a simple {description} of linear confinement in quenched QCD in terms of a gluon mass function describing the non-perturbative coupling.

There are several issues which have arisen in our investigation that merit attention and 
which we next recall.

{Confinement can be described by a one-gluon-exchange picture if some of the long-distance physics is folded in an effectively generated gluon mass. In that context} asymptotic freedom of the coupling and phenomenology require that the gluon mass function goes rapidly to zero, faster than $1/q^2$. This result is in complete agreement with lattice QCD and the SDE solutions. 

{ In that context, a Gribov  type singularity is the least requirement for linear confinement. This IR singular behavior is an {\it ad hoc} assumption, which many phenomenological studies of the IR coupling do not support \cite{Deur:2016tte}, and implies that the SDE formulation of the gluon mass function must be restricted to $m_0= \Lambda$}. If $m_0 > \Lambda$,  linear confinement will soften to a Yukawa type behavior. {In the present non-relativistic dynamical scheme this type of potentials are non confining, however this is not so in other formulations \cite{Brodsky:2014yha}.}

A main result of our analysis is that we are led to Cornell type potentials in the region relevant to describe the spectra
($0.1-4$ fm). In this way we {support} the phenomenological success associated with Cornell type potential as a 
consequence of QCD.

 Our fit of the spectra  is excellent. Note that we have used only one free parameter, which comes out at a very reasonable value ($\Lambda \sim 320$ MeV). Moreover, we justify the slight difference between the bottomonium and charmonium potentials in terms of the number of flavors entering the coupling.

Therefore we have shown that by incorporating the infrared Gribov singularity in a manner that respects the behavior of the massive SDE coupling  with fixed parameters  at large $q^2$ we obtain an excellent potential capable of reproducing the heavy hadron spectrum with only one parameter, $\Lambda$, which moreover comes at a wishful value $\sim 300$ GeV {close to the value determined from other phenomenology in the $\overline{MS}$ scheme}.  Moreover, the flavor dependence of the SDE coupling controls the flavor dependence of the potential. While our potential is very close to the Cornell potential in the physical range it differs from it outside the range and has a perfect asymptotic QCD behavior. Moreover this coupling not only can be used to explain the spectrum but can be used in many other processes where we are dealing with large to moderate energies.

We  conclude by stating that we have found an explanation for the phenomenological successful potentials in terms of a non-perturbative effective mechanism describing the strong coupling. This mechanism is  defined  by means of a gluon mass function with properties closely related to lattice QCD and SDE studies to which we have incorporated in a gentle way the Gribov singularity.  We have supported the successful Cornell type potentials as arising from specific mechanisms in non-perturbative QCD.

\begin{acknowledgments}
\noindent We acknowledge useful comments by G. Cveti\v{c}. This work has been supported in part by MINECO (Spain) Grants. Nos. FPA2013-47443-C2-1-P and
FPA2014-53631-C2-1-P, GVA-PROMETEOII/2014/066, SEV-2014-0398  and CA also by CONICYT ``Becas
Chile'' Grant No.74150052 .
\end{acknowledgments}


\begin{thebibliography}{99}                                                                                               %


\bibitem {Fritzsch:1973pi}H.~Fritzsch, M.~Gell-Mann and H.~Leutwyler,
Phys.\ Lett.\ B \textbf{47}, 365 (1973).




\bibitem {Dokshitzer:1977sg}Y.~L.~Dokshitzer,
Sov.\ Phys.\ JETP \textbf{46} (1977) 641 [Zh.\ Eksp.\ Teor.\ Fiz.\ \textbf{73}
(1977) 1216].




\bibitem {Gribov:1972ri}V.~N.~Gribov and L.~N.~Lipatov,
Sov.\ J.\ Nucl.\ Phys.\ \textbf{15} (1972) 438 [Yad.\ Fiz.\ \textbf{15} (1972)
781].




\bibitem {Altarelli:1977zs}G.~Altarelli and G.~Parisi,
Nucl.\ Phys.\ B \textbf{126} (1977) 298.




\bibitem {Gross:1973id}D.~J.~Gross and F.~Wilczek,
Phys.\ Rev.\ Lett.\ \textbf{30}, 1343 (1973).




\bibitem {Politzer:1973fx}H.~D.~Politzer,
Phys.\ Rev.\ Lett.\ \textbf{30}, 1346 (1973).




\bibitem {Wilson:1974sk}K.~G.~Wilson,
Phys.\ Rev.\ D \textbf{10} (1974) 2445.




\bibitem {Hagler:2011zz}P.~Hagler,
Prog.\ Theor.\ Phys.\ Suppl.\ \textbf{187} (2011) 221.




\bibitem {Bali:2000gf}G.~S.~Bali,
Phys.\ Rept.\ \textbf{343}, 1 (2001) [hep-ph/0001312].




\bibitem {Eichten:1978tg}E.~Eichten, K.~Gottfried, T.~Kinoshita, K.~D.~Lane
and T.~M.~Yan,
Phys.\ Rev.\ D \textbf{17}, 3090 (1978) [Phys.\ Rev.\ D \textbf{21}, 313
(1980)].




\bibitem {Quigg:1979vr}C.~Quigg and J.~L.~Rosner,
Phys.\ Rept.\ \textbf{56}, 167 (1979).




\bibitem {GI85}S. Godfrey and N. Isgur, Phys. Rev. D \textbf{32}, 189 (1985).



\bibitem {Kochelev:1985de}N.~I.~Kochelev,
Sov.\ J.\ Nucl.\ Phys.\ \textbf{41} (1985) 291 [Yad.\ Fiz.\ \textbf{41} (1985)
456].




\bibitem {Brambilla:1999ja}N.~Brambilla and A.~Vairo,
In *Newport News 1998, Strong interactions at low and intermediate energies*
151-220 [hep-ph/9904330].

\bibitem{Nesterenko:2001st}
  A.~V.~Nesterenko,
  Phys.\ Rev.\ D {\bf 64} (2001) 116009
  [hep-ph/0102124].
  
\bibitem{Nesterenko:2003xb}
  A.~V.~Nesterenko,
  Int.\ J.\ Mod.\ Phys.\ A {\bf 18} (2003) 5475
  [hep-ph/0308288].

  



\bibitem {Cornwall:1982zr}J.~M.~Cornwall,
Phys.\ Rev.\ D \textbf{26}, 1453 (1982).




\bibitem {Aguilar:2006gr}A.~C.~Aguilar and J.~Papavassiliou,
JHEP \textbf{0612}, 012 (2006).

\bibitem{Erlich:2005qh}
  J.~Erlich, E.~Katz, D.~T.~Son and M.~A.~Stephanov,
  Phys.\ Rev.\ Lett.\  {\bf 95} (2005) 261602
  doi:10.1103/PhysRevLett.95.261602
  [hep-ph/0501128].
  
\bibitem{Brodsky:2014yha}
  S.~J.~Brodsky, G.~F.~de Teramond, H.~G.~Dosch and J.~Erlich,
  Phys.\ Rept.\  {\bf 584} (2015) 1
  doi:10.1016/j.physrep.2015.05.001
  [arXiv:1407.8131 [hep-ph]].
  


\bibitem{Aguilar:2009nf}
  A.~C.~Aguilar, D.~Binosi, J.~Papavassiliou and J.~Rodriguez-Quintero,
  Phys.\ Rev.\ D {\bf 80} (2009) 085018
  [arXiv:0906.2633 [hep-ph]].
  
\bibitem{Ayala:2012pb}
  A.~Ayala, A.~Bashir, D.~Binosi, M.~Cristoforetti and J.~Rodriguez-Quintero,
  Phys.\ Rev.\ D {\bf 86} (2012) 074512
  doi:10.1103/PhysRevD.86.074512
  [arXiv:1208.0795 [hep-ph]].
  
\bibitem{Aguilar:2013hoa}
  A.~C.~Aguilar, D.~Binosi and J.~Papavassiliou,
  Phys.\ Rev.\ D {\bf 88} (2013) 074010
  doi:10.1103/PhysRevD.88.074010
  [arXiv:1304.5936 [hep-ph]].

\bibitem {Gonzalez:2011zc} P.~Gonzalez, V.~Mathieu and V.~Vento,
Phys.\ Rev.\ D \textbf{84}, 114008 (2011) [arXiv:1108.2347 [hep-ph]].



\bibitem {Greensite:2003xf}J.~Greensite, S.~Olejnik,
Phys.\ Rev.\ \textbf{D67 } (2003) 094503. [hep-lat/0302018].







\bibitem {Vento:2012wp}V.~Vento,
Eur.\ Phys.\ J.\ A \textbf{49} (2013) 71 [arXiv:1205.2002 [hep-ph]].






\bibitem {Richardson:1978bt}J.~L.~Richardson,
Phys.\ Lett.\ B \textbf{82}, 272 (1979).

\bibitem{Buchmuller:1980su}
  W.~Buchmuller and S.~H.~H.~Tye,
  Phys.\ Rev.\ D {\bf 24} (1981) 132.

\bibitem {Nesterenko:1999np}A.~V.~Nesterenko,
Phys.\ Rev.\ D \textbf{62}, 094028 (2000) [hep-ph/9912351].

\bibitem{Epele:2001ic}
  L.~N.~Epele, H.~Fanchiotti, C.~A.~Garcia Canal and M.~Marucho,
  Phys.\ Lett.\ B {\bf 523} (2001) 102
  [hep-ph/0103186].

\bibitem{Luna:2010tp}
  E.~G.~S.~Luna, A.~L.~dos Santos and A.~A.~Natale,
  Phys.\ Lett.\ B {\bf 698} (2011) 52
  [arXiv:1012.4443 [hep-ph]].


\bibitem{Gomez:2015tqa}
  J.~D.~Gomez and A.~A.~Natale,
  arXiv:1509.04798 [hep-ph].



\bibitem {Gribov:1999ui}V.~N.~Gribov,
Eur.\ Phys.\ J.\ C \textbf{10} (1999) 91 [hep-ph/9902279].




\bibitem {Simonov:1999qj}Y.~A.~Simonov,
hep-ph/9911237.




\bibitem {Yndurain:2000yq}F.~J.~Yndurain,
Nucl.\ Phys.\ Proc.\ Suppl.\ \textbf{93} (2001) 196 [hep-ph/0008007].

\bibitem{Deur:2016tte}
  A.~Deur, S.~J.~Brodsky and G.~F.~de Teramond,
  Prog.\ Part.\ Nucl.\ Phys.\  {\bf 90} (2016) 1
  doi:10.1016/j.ppnp.2016.04.003
  [arXiv:1604.08082 [hep-ph]].



\bibitem {Aguilar:2014tka}A.~C.~Aguilar, D.~Binosi and J.~Papavassiliou,
Phys.\ Rev.\ D \textbf{89}, 085032 (2014) [arXiv:1401.3631 [hep-ph]].



\bibitem{ITEP} 
  M.~A.~Shifman, A.~I.~Vainshtein and V.~I.~Zakharov,
  Nucl.\ Phys.\ B {\bf 147}, 385 (1979).






\bibitem {PDG14}K. A. Olive \textit{et al.} [Particle Data Group (PDG)], Chin.
Phys. C \textbf{38}, 090001 (2014).



\bibitem {Ynd95}S. Titard and F. J. Yndur\'{a}in\textit{, }Phys. Lett. B
\textbf{351}, 541 (1995); Phys. Rev. D \textbf{51}, 6348 (1995).



\bibitem {Bad99}A. M. Badalian and V. L. Morgunov, Phys. Rev. D \textbf{60},
116008 (1999).

\bibitem{Yndurain:1999ui}
  F.~J.~Yndurain,
  The theory of quark and gluon interactions,
  Berlin, Germany: Springer (2006).

\bibitem{Kataev:2001kk}
  A.~L.~Kataev, G.~Parente and A.~V.~Sidorov,
  Phys.\ Part.\ Nucl.\  {\bf 34} (2003) 20
   [Fiz.\ Elem.\ Chast.\ Atom.\ Yadra {\bf 34} (2003) 43]
   [Phys.\ Part.\ Nucl.\  {\bf 38} (2007) 6,  827]
  [hep-ph/0106221].


\bibitem{Kataev:2015yha}
  A.~L.~Kataev and V.~S.~Molokoedov,
  Phys.\ Rev.\ D {\bf 92},  054008
  [arXiv:1507.03547 [hep-ph]].


\end{thebibliography}
\end{document}